\documentclass[10pt,lettersize,journal]{IEEEtran}
\usepackage{amsmath,amsfonts}
\usepackage{algorithmic}
\usepackage{algorithm}
\usepackage{graphicx}
\usepackage{textcomp}
\usepackage{xcolor}
\usepackage[caption=false,font=normalsize,labelfont=sf,textfont=sf]{subfig}
\usepackage{textcomp}
\usepackage{stfloats}
\usepackage{verbatim}
\usepackage{cite}
\usepackage{tabularx,booktabs}
\usepackage{multirow}
\usepackage{multicol}
\usepackage{enumerate}
\usepackage{enumitem}
\usepackage{ragged2e}
\PassOptionsToPackage{hyphens}{url} 
\usepackage{hyperref}
\begin{document}

\title{Utilizing Priors in Sampling-based Cost Minimization}

\author{
    Yuan-Yao Lou,
    Jonathan Spencer,
    Kwang Taik Kim,
    Mung Chiang
    
}



\maketitle


\section{Problem Statement} \label{sec:problem_statement}

We consider an autonomous vehicle (AV) agent performing a long-term cost-minimization problem in the elapsed time $T$ over sequences of states $s_{1:T}$ and actions $a_{1:T}$ for some fixed, known (though potentially learned) cost function $C(s,a)$, approximate system dynamics $\mathcal{P}$, and distribution over initial states $d_0$. The goal is to minimize the expected cost-to-go of the driving trajectory $\tau = s_1, a_1, ..., s_T, a_T$ from the initial state. When T is very large, model predictive control (MPC; sometimes called receding horizon control) can solve the cost-minimization problem over a shorter horizon $H \ll T$. Specifically, MPC executes the first action of the short horizon minimization, then successively minimizes and executes subsequent actions based on the updated state. With the MPC policy, the total cost of the trajectory is
\begin{align} \label{eqn:cost}
    & C(\tau) = \sum_{t=1}^{T} \mathbb{E}_{s_t \sim P} [C(s_t,a_t)], \nonumber \\
    & \text{where} \ a_t \in \underset{a_{t:t+H-1}}{\arg\min} \ \mathbb{E}_{s_h \sim P} \Bigg[
    \sum_{h=t}^{t+H-1} C(s_h, a_h)\Bigg].
\end{align}

Given that the horizon $H$ is much shorter, we can simplify the computation by sampling motion primitives from a fixed dictionary $D_\alpha = \{\alpha^{(1)},...,\alpha^{(i)},...,\alpha^{(K)}\}$ with size $K$ where each primitive is a valid sequence of $H$ actions $\alpha^{(i)} = a_1^{(i)}\, ..., a_H^{(i)}$. Generally, the primitives are sampled from a distribution, which may potentially depend on the state $\alpha^{(i)} \sim p(\alpha|s)$, but in the simplest case $p(\alpha)$ is just the uniform distribution over motion primitives. For a deterministic system model $\mathcal{P}$, a motion primitive $\alpha^{(i)}$ produces a unique state sequence $s_{t:t+H}^{(i)}$, and the associated cost is $C(\alpha^{(i)}|s_t) = \sum_{h=0} ^{H-1} C(s_{h+t}^{(i)},a_h^{(i)})$. Using the primitive library, the choice of $a_t$ from (\ref{eqn:cost}) becomes
\begin{align} \label{eqn:action}
a_t=a_0^{(i)} \in \underset{\alpha^{(i)} \in D_\alpha}{\arg\min}~ C(\alpha^{(i)}|s_t).
\end{align}

\section{Cost-optimal prior trajectory library} \label{sec:system-framework}

We save the selected motion primitive $\alpha^{(i)}$ at each state $s_{t}$ along the cost-optimal prior trajectories $\tau$ that arrive at the destination successfully. In this work, a cost-optimal trajectory is stored if it has the lowest cost among all the sampling trajectories. Due to the computation limitation, for every data collection stage, we exclude the top 10\% of historical data with the highest and lowest cost-to-go and save the others into the cost-optimal trajectory library. In addition to computation and storage constraints, such a data-pruning approach based on statistical analysis is well-known in machine learning.

The cost-optimal prior data is stored at a set of anchor points $A$ by a state mapping function $f_{\text{mapping}}(s_i) = A_i$. Since an anchor point can contain several historical data if different states belong to the same anchor point, the cost-optimal prior library can be represented as a dictionary
$$D'_\textit{prior} = \{ (A, h)_i\ \lvert\ h = \{\alpha_k^{(j)}\} \},$$
where $h$ is a set of prior motion-primitive. For each anchor point with its stored prior data, we form an empirical distribution by accumulating the number of occurrences according to the stored prior data. The sampling probability of each trajectory is then calculated by dividing the number of occurrences by the total sample size. $$p(\alpha^{(j)} \lvert A_i \in D'_\textit{prior}) = \frac{\sum_{\alpha_{i'} \in h_i}  1[\alpha_{i'}=j]}{|h_i|}.$$
Given that the motion primitive sampling is based on the state $s$ and the storage constraint of cost-optimal prior library $D'_\textit{prior}$, uniform sampling is adopted if a state falls out of the library.
$$
\begin{cases}
    \alpha^{(i)} \sim P_\textit{optimal}, \text{if } f_\textit{mapping}(s) \in D'_\textit{prior}, \\
    \alpha^{(i)} \sim p(\alpha) = P_\textit{uniform}, \text{otherwise},
\end{cases}
$$
where $P_\textit{optimal}$ is the sampling probability mass function (PMF) from the cost-optimal motion-primitive library, and $P_\textit{uniform}$ is the sampling PMF from the uniform distribution. Moreover, considering that anchor points might not contain enough historical priors if data collection stages just begin, the AV providers might perform parameter sweeping on $\beta$ to decide the trust level on the cost-optimal prior library for sampling at different times.
$$
\begin{cases}
    P_\textit{optimal} = P_U(u) \cdot (1 - \beta) + P_E(e) \cdot \beta, \\
    P_\textit{optimal} = P_\textit{uniform}, \text{when } \beta = 0.
\end{cases}
$$

Considering that the several points $D_{\text{points}}$ can be sampled at each anchor point from the motion-primitive library in the space $X$, the spatial relation can be written as $D_{\text{points}} \in D_{\alpha} \subset X_{\text{space}}$. Then the average performance (cost-to-go) is given by:
$$
\mu_\textit{avg} = \frac{1}{\lvert D_{\alpha} \lvert} \sum_{\alpha^{(i)} \in D_{\alpha}} C(\alpha^{(i)} \lvert x).
$$
For $M$ driving iterations with $N \ll \lvert D_{\alpha} \lvert$ samplings per moving step:
\begin{enumerate}
    \item   For the i-th iteration that the agent enters an anchor point $x$ without historical data, the probability of getting a trajectory with a cost-to-go lower or higher than the mean value can be described as:
    $$
    \begin{cases}
        p_x (C(\alpha^{(i)} \lvert x) \leq \mu) = P_{\mu^{-}} = \frac{\lvert D_{-} \lvert}{\lvert D_{\alpha} \lvert}, \\
        p_x (C(\alpha^{(i)} \lvert x) > \mu) = P_{\mu^{+}},
    \end{cases}
    $$
    where $D_{-}$ is a set of points on the trajectories with better performance compared to the $\mu$. Thus, the lowest possibility to outperform average cost-to-go with consecutive $N$ sampling is defined as:
    $$\delta_{1} = 1 - (1 - P_{\mu^{-}})^N.$$
    
    \item   For the m-th iteration that the agent enters an anchor point $x$ stored with several historical data, the probability of getting a better trajectory is expected to be higher than $\delta_{1}$. Assuming several points $m_{-}$ with better performance are collected, the trajectory sampling now belongs to the cost-optimal library, $\alpha^{(j)} \in D'_{\textit{prior}}$. Then, the probability of getting a trajectory with a cost-to-go lower than the mean value can be described as:
    $$
    p_x (C(\alpha^{(j)} \lvert x) \leq \mu) = P'_{\mu^{-}} =\frac{m_{{-}}}{m},
    $$
    $$
    \delta_{2} = 1 - (1 - P'_{\mu^{-}})^N \geq \delta_{1} = 1 - (1 - P_{\mu^{-}})^N,
    $$
    $$
    \frac{m_{{-}}}{m} \geq \frac{\lvert D_{-} \lvert}{\lvert D_{\alpha} \lvert},
    $$
    The probability of getting a trajectory with lower cost by fully relying on empirical distribution depends on the amount of driving iterations $m$, the size of sampling points $D$, and the number of collected data points $m_{-}$.
\end{enumerate}

{\appendix[State-Action Cost Function]\label{sec:appendix}

To evaluate the set of candidate motion primitives and the corresponding predicted trajectories, the state-action cost function is a linear feature weighting $C(s, a) = w^Tf(s, a)$. The feature weights $w$ are highly sensitive and hand-tuned so that the agent approximately tracks the reference path while producing a desirable trajectory in terms of smoothness and efficiency. One possible form of the features $f(s, a)$ that make up the cost function consists of the following real-value functions, written as:

$$ f(s,a) = \begin{bmatrix}DistFrechet(s_t) \\ SteeringMagnitude(a^{(i)}) \\ DistDestination(s_t) \\ DistObstacle(s_t) \\ CollisionDetection(s_t) \\ Disturbance(s_t) \end{bmatrix}$$

\begin{itemize}[leftmargin=*]
    \item   \textbf{Fréchet distance}:
    The proximity between the predicted trajectories and the global reference path is computed using Fréchet distance \cite{alt1995computing}.\footnote{Fr\'echet distance is a distance metric commonly used to compare trajectories of potentially uneven length. Informally, given a person walking along one trajectory and a dog following the other without either backtracking, the Fr\'echet distance is the length of the shortest possible leash for both to make it from start to finish.} Given that the reference path is feasible but not an optimal path, this feature gives the agent the flexibility to deviate. $$ DistFrechet(s_t) = \text{SimilarityCurve}(\cup^{H-1}_{h=0} s^{(i)}_{t+h}, \text{ref\_path}) $$

    \item   \textbf{Magnitude of steering controls}:
    The magnitude of steering controls is calculated by computing the difference in the action sequence of the predicted trajectory and summing their absolute values, which penalizes abrupt steering. $$SteeringMagnitude(a^{(i)}) = \Sigma^{H-1}_{h=1} \lvert a^{(i)}_{h} - a^{(i)}_{h-1} \lvert$$

    \item   \textbf{Distance from the destination}:
    We use average Euclidian distance to the goal when the agent has a clear line of sight and a constant maximum feature value is applied otherwise. $$DistDestination(s_t) = \Sigma^{H-1}_{h=0} \lvert\lvert s^{(i)}_{t+h} - \text{dest} \lvert\lvert_2$$

    \item   \textbf{Distance from the obstacles}:
    This feature uses average inverse Euclidian distance to obstacles so that when the trajectory passes near obstacles the feature is large. Since the reference path supplied by the global planner is oblivious to obstacle proximity, it may in some cases pass unreasonably close. This feature balances the desire to coarsely track the reference while ensuring collision avoidance. $$DistObstacle(s_t) = \frac{1}{\Sigma^{H-1}_{h=0} \Bigl[ \Sigma_{i} \lvert\lvert s^{(i)}_{t+h} - \text{obst}_i \lvert\lvert_2 \Bigr]}$$

    \item   \textbf{Obstacle collision detection}:
    The weight for this feature is very large so any intersection with an obstacle suffers a large cost. When all of the predicted trajectories collide with obstacles, we label that state as a \textit{blind spot}. $$CollisionDetection(s_t) = \Sigma^{H-1}_{h=0} \Bigl[ \Sigma_j\ s^{(i)}_{t+h} \cap \text{obst}_j \Bigr]$$

    \item   \textbf{High-cost regions}:
    We assume that sometimes the agent encounters regions (for example icy roads or mud areas) which are undetectable by the agent but detectable using sensors on the edge node. These regions affect the motion model unpredictably, so we assign them high costs to encourage the agent to avoid them where possible. The edge nodes compute this as the average over the trajectory of the overlap indicator. Since the agents cannot sense this feature, they must pessimistically assume the high-cost region covers the entire map. $$Disturbance(s_t) = \Sigma^{H-1}_{h=0} \Bigl[ \Sigma_j\ s^{(i)}_{t+h} \cap \text{region}_j \Bigr]$$
\end{itemize}

\bibliographystyle{IEEEtran}
\bibliography{reference}

\begin{thebibliography}{1}
\providecommand{\url}[1]{#1}
\csname url@samestyle\endcsname
\providecommand{\newblock}{\relax}
\providecommand{\bibinfo}[2]{#2}
\providecommand{\BIBentrySTDinterwordspacing}{\spaceskip=0pt\relax}
\providecommand{\BIBentryALTinterwordstretchfactor}{4}
\providecommand{\BIBentryALTinterwordspacing}{\spaceskip=\fontdimen2\font plus
\BIBentryALTinterwordstretchfactor\fontdimen3\font minus
  \fontdimen4\font\relax}
\providecommand{\BIBforeignlanguage}[2]{{%
\expandafter\ifx\csname l@#1\endcsname\relax
\typeout{** WARNING: IEEEtran.bst: No hyphenation pattern has been}%
\typeout{** loaded for the language `#1'. Using the pattern for}%
\typeout{** the default language instead.}%
\else
\language=\csname l@#1\endcsname
\fi
#2}}
\providecommand{\BIBdecl}{\relax}
\BIBdecl

\bibitem{alt1995computing}
H.~Alt and M.~Godau, ``{Computing the {Fr\'e}chet distance between two
  polygonal curves},'' \emph{International Journal of Computational Geometry \&
  Applications}, vol.~5, pp. 75--91, 1995.

\end{thebibliography}

\end{document}